\documentclass[aps,prl, amsmath, reprint, amssymb, superscriptaddress]{revtex4-2}
\usepackage{graphicx}% include figure files
\usepackage{dcolumn}% Align table columns on decimal point
\usepackage{bm}% bold math
\usepackage{siunitx}
\usepackage{hyperref}

\hypersetup{
	breaklinks=true,   % splits links across lines
	colorlinks=true,   % displays links as colored text instead of blocks
	pdfusetitle=true,  % \title and \author values into pdf metadata
}

\begin{document}

%\preprint{}

\title{Trapping Ion Coulomb Crystals in an Optical Lattice}


\author{Daniel Hoenig}
\affiliation{Albert-Ludwigs-Universität Freiburg, Physikalisches Institut, 79104 Freiburg, Germany}
%\email[]{daniel.hoenig@physik.uni-freiburg.de}
%\thanks{}
\author{Fabian Thielemann}
\affiliation{Albert-Ludwigs-Universität Freiburg, Physikalisches Institut, 79104 Freiburg, Germany}
\author{Leon Karpa}
\affiliation{Albert-Ludwigs-Universität Freiburg, Physikalisches Institut, 79104 Freiburg, Germany}
\affiliation{Leibniz Universität Hannover, Institut für Quantenoptik, 30167 Hannover, Germany}

\author{Thomas Walker}
\affiliation{Albert-Ludwigs-Universität Freiburg, Physikalisches Institut, 79104 Freiburg, Germany}
\author{Amir Mohammadi}
\affiliation{Albert-Ludwigs-Universität Freiburg, Physikalisches Institut, 79104 Freiburg, Germany}
\author{Tobias Schaetz}
\email[]{tobias.schaetz@physik.uni-freiburg.de}
\affiliation{Albert-Ludwigs-Universität Freiburg, Physikalisches Institut, 79104 Freiburg, Germany}

\date{\today}

\begin{abstract}
	
We report the optical trapping of multiple ions localized at individual lattice sites of a one-dimensional optical lattice.
We observe a fivefold increase in robustness against axial DC-electric fields and an increase of the axial eigenfrequency by two orders of magnitude compared to an optical dipole trap without interference but similar intensity.
Our findings motivate an alternative pathway to extend arrays of trapped ions in size and dimension, enabling quantum simulations with particles interacting at long range.
\end{abstract}

%\keywords{}

\maketitle

%%% ------------------------INTRODUCTION----------------------------------------------------------------------------
Analog quantum simulation - exploiting well controlled quantum systems to experimentally simulate phenomena in nature, which are otherwise hard to access - has seen great success with the emergence of several well controlled experimental platforms \cite{feynmanSimulatingPhysicsComputers1982,ciracGoalsOpportunitiesQuantum2012,schaetzFocusQuantumSimulation2013,georgescuQuantumSimulation2014, altmanQuantumSimulatorsArchitectures2021}.
However, it remains an outstanding objective to explore the regime beyond efficient numerical tractability, for example by addressing the class of problems incorporating interaction at long range while exceeding one dimension (1D) \cite{ciracGoalsOpportunitiesQuantum2012}.
Promising approaches include extending the interaction range between optically trapped neutral particles by employing strongly dipolar (Rydberg) atoms \cite{baierExtendedBoseHubbardModels2016,wuConciseReviewRydberg2021} or molecules \cite{blackmoreUltracoldMoleculesQuantum2018} or using multi-species atomic ensembles coupled to cavities \cite{arguello-luengoAnalogueQuantumChemistry2019}.

Trapping ion Coulomb crystals (CCs) in radiofrequency (rf) traps allows the direct exploitation of the long range Coulomb interaction \cite{thompsonIonCoulombCrystals2015,drewsenIonCoulombCrystals2015}.
Experiments with linear CCs have led to seminal results (see \cite{blattQuantumSimulationsTrapped2012,schneiderExperimentalQuantumSimulations2012,monroeProgrammableQuantumSimulations2021} and references therein).
Yet, extending the approach to two- or three-dimensional CCs in linear rf traps enforces a spatial displacement of the ions from the rf node, causing an rf-driven motion, the so called micromotion.
Aligning the orientation of the interaction perpendicular to the micromotion, could substantially reduce the impact for some applications \cite{donofrioRadialTwoDimensionalIon2021,wangQuantumComputationMicromotion2015,kiesenhoferControllingTwodimensionalCoulomb2023}.
Still, already in few ion CCs, the kinetic energy of the synchronized micromotion exceeds the thermal energy by several orders of magnitude, similar to the approach in Penning traps \cite{drewsenIonCoulombCrystals2015}.
Trapping ions in arrays of individual rf surface traps mitigates the micromotion in higher dimensions, and permits single site control \cite{schaetzScalableQuantumSimulations2007,chiaveriniLaserlessTrappedionQuantum2008,sterlingFabricationOperationTwodimensional2014, mielenzArraysIndividuallyControlled2016, warringTrappedIonArchitecture2020, holz2DLinearTrap2020}. 
However, extending the size of the arrays, while maintaining sufficiently small electrode structures, remains a challenge.
Hybrid traps, replacing the axial confinement in rf traps by a 1D optical lattice were successfully employed to axially pin ions by the light field \cite{katoriAnomalousDynamicsSingle1997,linnetPinningIonIntracavity2012, karpaSuppressionIonTransport2013, laupretreControllingPotentialLandscape2019}.
This has been used to simulate friction in 1D on an atomic scale \cite{cetinaOnedimensionalArrayIon2013,bylinskiiTuningFrictionAtombyatom2015,gangloffVelocityTuningFriction2015,bylinskiiObservationAubrytypeTransition2016,gangloffKinksNanofrictionStructural2020} and was proposed as platform to study structural phase-transitions in 2D \cite{horakOpticallyInducedStructural2012}.

All-optical trapping of CCs, i.e. in absence of any rf fields offers a possibility to extend arrays of ions in size and dimension.
Firstly, micromotion is negligible for optically trapped ions \cite{cormickTrappingIonsLasers2011}. 
Secondly optical fields, exploiting holography and interference \cite{barredoSyntheticThreedimensionalAtomic2018}, allow realizing and dynamically controlling, close to arbitrary potential landscapes, such as multidimensional microtrap arrays on the nanometer scale. 
Finally all-optical trapping allows for joint confinement of atoms and ions \cite{schmidtOpticalTrapsSympathetic2020,peregoElectroOpticalIonTrap2020} potentially extending the quantum simulation toolbox to arrays of ions and (Rydberg) atoms \cite{lesanovskyTrapassistedCreationGiant2009,schaetzTrappingIonsAtoms2017,mukherjeeChargeDynamicsMolecular2019}.
All-optical trapping of a single ion \cite{schneiderOpticalTrappingIon2010,huberFaroffresonanceOpticalTrap2014,lambrechtLongLifetimesEffective2017} as well as CCs \cite{schmidtOpticalTrappingIon2018,weckesserTrappingShapingIsolating2021} has been demonstrated for single-beam optical dipole traps (ODTs).
Additionally a single ion has been confined in a near-resonant optical lattice \cite{enderleinSingleIonsTrapped2012}. 
In this letter, we demonstrate optical trapping of multiple ions in a far-detuned optical lattice, realizing single-site localization at individual lattice sites.
We compare the eigenfrequencies of the ions in the lattice to those of ions in a non-interfering ODT and observe an increase of the axial eigenfrequency by two orders of magnitude.

\begin{figure}
	\includegraphics{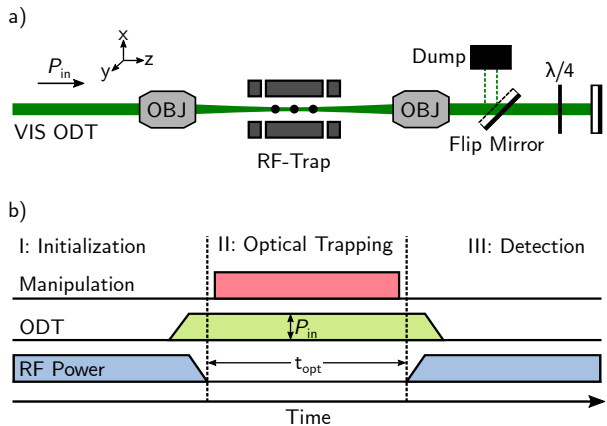}%
	\caption{\label{fig:setup}(color online) Schematic representation of the experiment. (a) Setup (not to scale):
		The setup consists of a linear rf trap with segmented DC blades and an optical dipole trap (ODT).
		The ODT is generated by a linear polarized beam ($\lambda = \qty{532}{\nano\metre}$) with power $P_\text{in}$, aligned along the common z axis.
		After the beam passes the trap, we either block it with a flip mirror, or retroreflect it, with the polarization of the retroreflection set with a quarter-wave plate.
		(b) Sketch of the experimental sequence: 
		We prepare a linear Coulomb crystal in the rf trap (I: Initialization). We transfer the crystal into one of the ODT configurations (II: Optical Trapping) and optionally act on the ions with additional fields. 
		After the optical trapping duration $t_\text{opt}$, we detect remaining ions in the rf trap (III: Detection).}
\end{figure}

%%% ---------------------------------------SETUP------------------------------------------------------------
A schematic representation of the experimental setup is given in Fig.\ref{fig:setup}(a).
We initially trap linear CCs of $^{138}\text{Ba}^{+}$ ions in a linear segmented rf trap (trap axis defines z axis) with secular frequencies $(\omega_\text{x}, \omega_\text{y}, \omega_\text{z}) \approx 2\pi \times(100,100,12)\ \rm{kHz}$ at driving frequency $\Omega_\text{rf} = 2\pi\times\qty{1.416}{\mega\hertz}$.
The axial secular frequency $\omega_\text{z}$ receives contributions from the rf field ($\approx 2\pi\times\qty{3}{\kHz}$) as well as from additional DC confinement supplied by the outer segments (endcaps) of the trap ($\omega_\text{z,DC} \approx 2\pi\times\qty{11.5}{\kilo\hertz}$).
The ions experience a magnetic field of approximately \qty{300}{\micro\tesla} at an angle of about \qty{75}{\degree} with respect to the z axis.
We generate an ODT with a linear polarized laser at wavelength $\lambda=\qty{532}{\nano\metre}$ and power $P_\text{in} \in \left[ 0,7 \right] \unit{\watt}$ aligned along the z axis.
We focus the incident beam with a first objective to a beam waist ($1/e^2$ of intensity) of $w_\text{in}\approx\qty{6.6}{\micro\metre}$ at the rf-node, where we position the CC's center of mass.
After the beam passes the vacuum chamber, we collimate it with a second objective, before optionally retroreflecting and refocusing it on the ion with $P_\text{ret} \approx 0.86 \times P_\text{in}$ and a beam waist matching $w_\text{in}$ within 10\%.
We realize three different ODT configurations, assisted by the axial DC-confinement:
(i) a single-beam-ODT, by blocking the incident beam with the flip mirror behind the chamber.
(ii) a lin$\perp$lin-ODT by rotating the polarization of the retroreflection by \qty{90}{\degree} with a quarter-wave plate.
(iii) a lin$\parallel$lin-ODT obtained by overlapping the incident beam with a parallel polarized retroreflection.
In this configuration the two beams interfere constructively, forming a 1D optical lattice.

%%% -----------------------------------------------------Methods---------------------------------------------------- 
Our optical trapping experiments consist of three phases (Fig.\ref{fig:setup}(b)):
In the initial phase, we load ions into the rf trap and laser cool them (D1-line) close to the Doppler temperature $T_\text{D} \approx \qty{400}{\micro\kelvin}$ in all motional degrees of freedom.
We monitor the ions by fluorescence imaging, to prepare isotope-pure CCs of $N\leq3$ ions in the $6S_{1/2}$-state as described in \cite{schmidtMassselectiveRemovalIons2020} and \cite{weckesserTrappingShapingIsolating2021}, and compensate stray fields to $E_\text{stray} \leq \qty{10}{\milli\volt\per\metre}$.
For the second phase, we transfer the CC from the rf trap into the ODT. To this end, we increase $P_\text{in}$ to its chosen value within $t_\text{ramp}\approx\qty{100}{\micro\second}$ and subsequently turn off the rf trap (ringdown time of \qty{32}{\micro\second}).
During the optical trapping duration $t_\text{opt}$, we optionally manipulate the ions with electric control fields.
In the final phase, we turn the rf trap on, switch the ODT off and detect remaining ions via fluorescence imaging.
We register a CC as optically trapped, if all ions are detected and repeat the protocol to derive the optical trapping probability $p_\text{opt}$.
As measure of statistical uncertainty for $p_\text{opt}$ we employ Wilson-Score 1-$\sigma$ intervals \cite{wilsonProbableInferenceLaw1927}.
%%% -------------------------------------RADIAL CONFINEMENT ------------------------------------------------------------

\begin{figure}
	\includegraphics{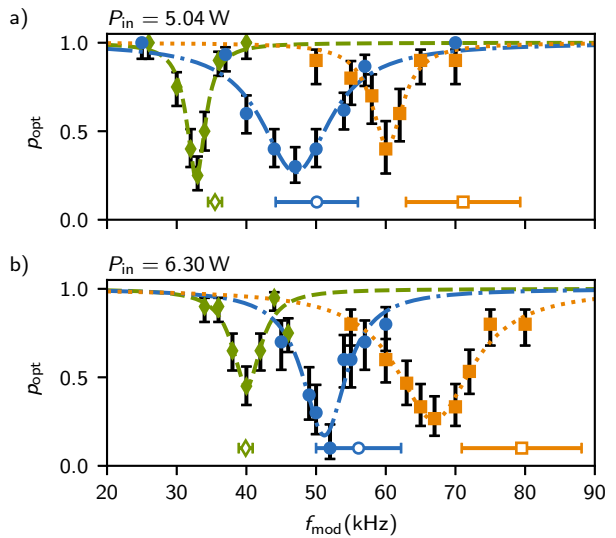}
	\caption{(color online)\label{fig:rad} Spectrometry of the radial modes of a single ion in different ODT configurations for a) \qty{5.04}{\watt} and b) \qty{6.30}{\watt}. Shown is $p_\text{opt}$ at different excitation frequencies $f_\text{mod}$ for the single-beam (green diamonds), the lin$\perp$lin (blue circles) and the lin$\parallel$lin-ODT (orange squares).
	Lines depict Lorentzian fits to the data. Open symbols show numerical estimations for the resonance frequencies (errorbars accounting for experimental uncertainties of model parameters).}
\end{figure}
In order to study the effect of the ODTs on the radial confinement of the ions, we investigate the radial motional eigenfrequencies $f_\text{rad,exp}$ of a single ion for the different configurations.
To derive $f_\text{rad,exp}$, we sinusodially modulate the voltage on one of the endcaps during $t_\text{opt} = \qty{1.5}{\milli\second}$ and measure $p_\text{opt}$ in dependence on the modulation frequency $f_\text{mod}$.
Tuning $f_\text{mod}$ close to resonance, we find a reduction of $p_\text{opt}$ due to coherent motional excitation.
We compare the radial eigenfrequencies for the three ODT configurations at $P_{\rm{in}}=\qtylist{5.04;6.30}{\watt}$ (Fig.\ref{fig:rad} (a) and (b), respectively).
The observed $f_\text{rad,exp}$ scale approximately with $\sqrt{P_\text{in}}$.
Additionally, the measurements show a significant increase in radial confinement from the single-beam- to the lin$\perp$lin- and lin$\parallel$lin-ODT at a given $P_\text{in}$.
We attribute the increase of $f_\text{rad,exp}$ in the lin$\perp$lin-ODT compared to the single-beam-ODT to the additional impact of $P_\text{ret}$, and 
the further increase in the lin$\parallel$lin-ODT to the interference of the beams.
As a comparison, we estimate radial frequencies $f_\text{rad,num}$, based on a harmonic approximation of a numerically modelled trap potential. 
While all $f_\text{rad,exp}$ agree with their related $f_\text{rad,num}$ to within \qty{1.5}{\sigma} \footnote{See Supplemental Material for a table listing experimentally and numerically obtained values for the resonance frequencies.},
they remain systematically below the $f_\text{rad,num}$ for the lin$\perp$lin and lin$\parallel$lin configurations.
We explain the deviation by non-Gaussian ODT-beams, as well as radial misalignment between the beams.
The radial anharmonicity of the trap potential leads to a further decrease of $f_\text{rad,exp}$ for increased ion kinetic energy $E_\text{kin}$.
The latter receives contributions from: (i) a non-adiabatic trap-transfer, inducing axial (radial) excitations, (ii) crossing of radial trap-instabilities during rf-rampdown \cite{takaiNonlinearResonanceEffects2007} and (iii) resonant axial excitation by the rf field during the transfer into the lin$\parallel$lin-ODT.
%%% ----------------------------------------------------------------------- DISPLACEMENT-----------------------------------------------------------------------------------------------------------------

\begin{figure}
	\includegraphics{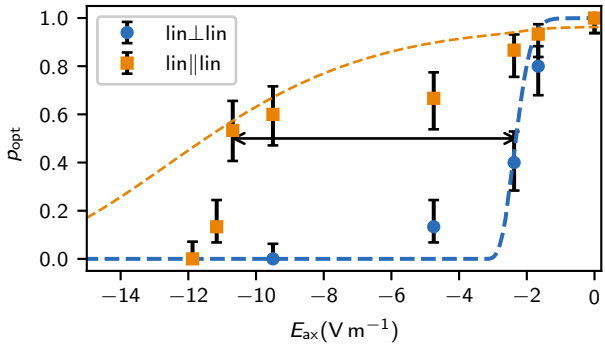}%
	\caption{(color online) \label{fig:disp} Optical trapping probability $p_\text{opt}$ for a single ion in the lin$\perp$lin- (blue circles) and lin$\parallel$lin-ODT (orange squares), exposed to an axial electric field $E_\text{ax}$. To reach comparable intensities in the two ODTs, we set $P_\text{in}$ to \qty{3.25}{\watt} and \qty{6.30}{\watt} respectively. As the black arrow indicates, we can apply an approximately five times larger $E_\text{ax}$ in the lin$\parallel$lin configuration, before $p_\text{opt}$ decreases to 50\%. The dashed blue curve shows the result of a simulation for ions with thermally distributed energy at temperature $T\approx 3 \times T_\text{D}$, neglecting any axial confinement by the light field. The orange dashed curve represents a simulated result for an initially optically single-site confined ion at $T \approx 6 \times T_\text{D}$.
	}
\end{figure}

To probe the axial confinement in the lin$\parallel$lin- and lin$\perp$lin-ODT, we measure $p_\text{opt}$ for a single ion in dependence on a homogeneous axial electric field $E_{\text{ax}}$ (Fig.\ref{fig:disp}). 
Anticipating an enhancement by interference, we compare the lin$\parallel$lin-ODT at $P_\text{in} =\qty{3.25}{\watt}$ to the lin$\perp$lin-ODT at $P_\text{in} =\qty{6.30}{\watt}$, to provide comparable peak intensities.
During $t_\text{opt}$, we increase $E_\text{ax}$ within \qty{200}{\micro\second} to its dedicated value and keep it for \qty{500}{\micro\second}. 
We find $p_\text{opt}$ enhanced for the lin$\parallel$lin configuration at $E_\text{ax}\geq\qty{2}{\volt\per\metre}$ allowing the application of a five times larger $E_\text{ax}$ before $p_\text{opt}$ decreases to $50\%$.
Note, that even though we apply more optical power (and comparable intensity) in the lin$\perp$lin configuration, it is significantly less robust against axial displacement fields.
We compare the measured $p_\text{opt}$ with the results of two simulations; one accounting for axial confinement by the light field, the other neglecting it.
In both cases, $E_\text{ax}$ displaces the ion axially, leading to reduced radial confinement.
For an ion, that is axially confined by the light field, $E_\text{ax}$ can additionally lead to an axial escape from its initial lattice site to a site with reduced radial confinement.
Accounting for these effects and assuming thermally distributed $E_\text{kin}$ with temperature T, we calculate $p_\text{opt}$ given $E_\text{ax}$, according to the (radial)-cutoff model \cite{schneiderInfluenceStaticElectric2012}.
For the lin$\perp$lin-ODT we find good agreement with the experiment for $T\approx3\times T_\text{D}$, without the inclusion of axial optical confinement.
We therefore do not find evidence for axial confinement by a polarization gradient, in principle possible for this configuration \cite{deutschQuantumstateControlOptical1998}.
To reproduce the enhanced $p_\text{opt}$ in the lin$\parallel$lin-ODT, however, we have to include axial optical confinement and derive $T \approx 6 \times T_\text{D}$.
We attribute the enhancement of the axial restoring force to the interference of the beams, confining the ion at a single lattice site.
The increase in $T$ suggests a further increased motional excitation during the transfer into the lin$\parallel$lin-ODT.
Apart from the imperfections discussed above, we suspect a residual misalignment of $E_\text{ax}$, causing a decrease of the radial trap depth with increased $E_\text{ax}$.
Additionally we expect a nonthermal energy distribution in the ODT, due to the energy increase during trap-transfer.

%%% -------------------------------------------------------------------------Axial Spectroscopy--------------------------------------------------------------------------------------------------------------
\begin{figure}
	\includegraphics{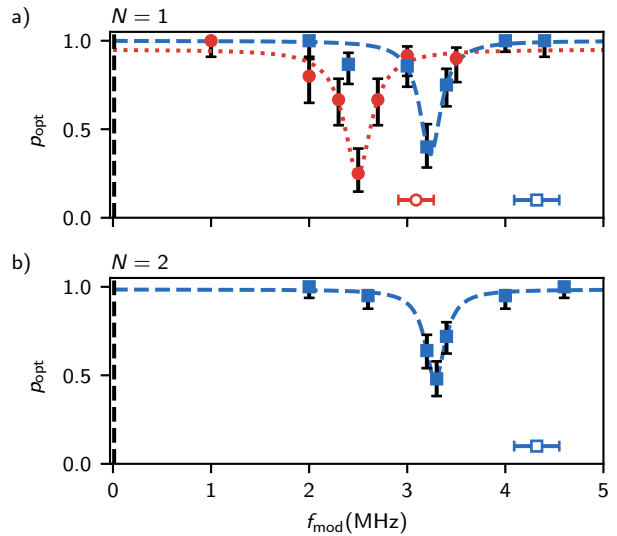}%
	\caption{\label{fig:ax}(color online) Spectrometry of the axial eigenfrequencies in the lin$\parallel$lin-trap for (a) a single ion and (b) a two ion crystal.
		 The measurements were performed at $P_{\rm{in}}=\qty{3.30}{\watt}$ (red circles) and $P_{\rm{in}}=\qty{6.30}{\watt}$ (blue squares). Shown is $p_\text{opt}$, in dependence on the frequency $f_\text{mod}$ of an axial driving field. The dashed and dotted lines show the result of a Lorentzian fit to the data. Open symbols represent an estimation of the expected frequencies based on numerical calculations. The observed frequencies are two orders of magnitude larger than the eigenfrequency measured for the single-beam-ODT, located at the dashed black line.
	 }
	 \end{figure}
 
 As an independent measure of the axial confinement in the lin$\parallel$lin-ODT, we derive the axial eigenfrequencies $f_\text{ax,exp}$ of a single ion for $P_{\text{in}}=\qty{3.30}{\watt} (\qty{6.30}{\watt})$ (Fig.\ref{fig:ax} (a)).
 During $t_\text{opt}$, we first apply a voltage offset on one of the endcaps to generate $E_\text{ax} \approx \qty{0.7}{\volt\per\metre} (\qty{1.1}{\volt\per\metre}) $, before additionally modulating the voltage on the opposite endcap at frequency $f_\text{mod}$ for \qty{500}{\micro\second}.
 Neither $E_\text{ax}$ nor the modulation alone substantially reduce $p_\text{opt}$.
 We attribute the need for the additional $E_\text{ax}$ to the anharmonicity of the axial potential, inhibiting sufficiently large motional amplitudes by excitation at $f_\text{mod}$.
 Similar to $f_\text{rad,exp}$, we observe a scaling of $f_\text{ax,exp}\sim\sqrt{P_\text{in}}$, evidencing that the measured confinement originates from the optical potential.
 Furthermore, the obtained $f_\text{ax,exp}$ are more than two orders of magnitude larger than the axial frequency observed in the single-beam-ODT (agreeing with pure DC-confinement).
 We repeat the measurement for $P_\text{in}=\qty{6.30}{\watt}$, $E_\text{ax}\approx\qty{0.7}{\volt\per\metre}$ and a CC of $N=2$ (Fig.\ref{fig:ax}(b)).
 The agreement of $f_\text{ax,exp}$ for $N=2$ and $N=1$ at identical $P_\text{in}$ shows, that the ions in the CC are initially axially confined within individual lattice sites.
 Using the same model as for $f_\text{rad,num}$, we derive numerical approximations $f_\text{ax,num}$.
 All $f_\text{ax,exp}$ remain at 80\% (76\%) of $f_\text{ax,num}$ for the \qty{3.30}{\watt} (\qty{6.30}{\watt}) measurement \cite{Note1} %(Table \ref{tab:results}).
 We explain the increased deviation of $f_\text{ax,exp}$ by a larger increase of $E_\text{kin}$ for axial compared to radial motion due to motional excitation during trap transfer.

%%% ----------------------------------------------------- CRYSTAL TRAPPING ----------------------------------------------------------------------------------------------------------------------------------

\begin{figure}
	\includegraphics{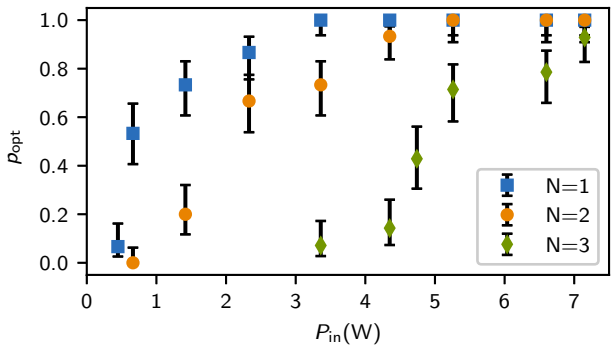}%
	\caption{\label{fig:crystal} (color online) Optical trapping probability $p_{\rm{opt}}$ for ion Coulomb crystals in the lin$\parallel$lin trap with one (blue squares), two (orange circles) and three ions (green diamonds) in dependence on the incident beam power $P_{\rm{in}}$. 
	We achieve trapping of CCs of up to $N=3$ ions with $p_\text{opt}$ close to unity.}
\end{figure}

Finally, we trap CCs with $N=1,2,3$ ions in the lin$\parallel$lin for $t_{\text{opt}} = \qty{500}{\micro\second}$ and measure the dependence of $p_\text{opt}$ on $P_\text{in}$ (Fig.\ref{fig:crystal}).
We achieve trapping of up to $N=3$ in the lin$\parallel$lin-trap with $p_\text{opt}$ close to one.
As $N$ increases, we need considerably more power, to reliably trap the CCs.
This is expected, as the ion for $N=1$ is situated at the rf node (axial and radial) and at the focus where the optical mode is smallest and cleanest while the ions for $N>1$ are generally not.
Additionally, the Coulomb repulsion between the ions reduces the trap depth (Resulting in an additional deconfinement of $\approx \qty{15}{\kilo\hertz}$ for the central ion at $N=3$) further lowering the expected $p_\text{opt}$. 
Finally, due to the larger rf field strength at increased axial distance to the rf node, the increase of $E_\text{kin}$ by resonant excitation during trap transfer gets enhanced for larger $N$.
An estimation of ion energies based on the radial-cutoff model yields temperatures on the order of $(5, 10, 20) \times T_D$ for $N=1,2,3$.

%%%----- OUTLOOK--------------------------------------------------------------
In this paper we have shown that we can trap CCs with up to three ions in an optical lattice.  
In our measurements we observed the effect of excess heating during trap transfer, currently limiting us in the size of trappable CCs.
By measuring the increased robustness of $p_\text{opt}$ against axial displacement and the axial eigenfrequencies of the ions in the lattice, we demonstrated that for $N\leq2$ the ions are single site confined at individual lattice sites even with increased kinetic energies.

One approach to mitigate the kinetic excitation during ODT loading, is an improved adiabatic transfer -- for example, via an intermittent single-beam-ODT, eventually building up the lattice with a second, individually controlled beam \cite{enderleinSingleIonsTrapped2012}.
An alternative is to in-situ ionize optically trapped neutral atoms, achieving small recoil energies by employing near-threshold photoionization or field ionization of Rydberg states \cite{engelObservationRydbergBlockade2018, dieterleInelasticCollisionDynamics2020}.
Additionally this would allow the inter-ion distance to be reduced, while easing the loading process for higher-dimensional ensembles in lattices \cite{schaetzTrappingIonsAtoms2017} or into optical-tweezer arrays with adjustable spacing.
To reach close to the motional ground state within individual lattice sites, we might exploit established sideband cooling schemes \cite{karpaSuppressionIonTransport2013}.
Alternatively, we could exploit sympathetic cooling via a bath of ultracold atoms \cite{schmidtOpticalTrapsSympathetic2020, weckesserObservationFeshbachResonances2021}. This might enable continuous operation, since the electronic degree of freedom is not involved.
However, even at reduced temperatures, an increased laser power might be required, to extend the ensemble.
This can be assisted by the use of optical cavities as demonstrated for optically trapped neutral atoms \cite{ritschColdAtomsCavitygenerated2013, hamiltonAtomInterferometryOptical2015} and hybrid ion traps\cite{linnetPinningIonIntracavity2012,karpaSuppressionIonTransport2013,laupretreControllingPotentialLandscape2019,bylinskiiTuningFrictionAtombyatom2015, gangloffVelocityTuningFriction2015,bylinskiiObservationAubrytypeTransition2016, gangloffKinksNanofrictionStructural2020}.
The ions could be imaged with fluorescence detection either in-situ while Raman cooling \cite{karpaSuppressionIonTransport2013} or free of effects induced by the ODT, by stroboscopically interleaving trapping and detection cycles.
Using the $6S_{1/2} \to 5D_{5/2}$ transition \cite{yumOpticalBariumIon2017} could then allow coherent control over the optical confinement, enabling the creation and study of coherent superpositions of structural crystal phases \cite{fishmanStructuralPhaseTransitions2008,retzkerDoubleWellPotentials2008,baltruschQuantumSuperpositionsCrystalline2011} and their entanglement.
Providing higher-dimensional lattices, filled with neutral atoms and ions, could allow the investigation of novel aspects of atom-ion interaction, e.g. involving coherent charge transfer between lattice sites hosting either ions or atoms \cite{lesanovskyTrapassistedCreationGiant2009,mukherjeeChargeDynamicsMolecular2019}, the latter in circular Rydberg-states \cite{patschFastAccurateCircularization2018}.
Yet, the Coulomb interaction at long range comes along with mutual repulsion, limiting the filling factor of the lattice. 
However, the stiff confinement within the accessible optical landscape might permit to address complex dynamics of interest, such as, the onset of Aharonov-Bohm physics via artificial gauge fields \cite{bermudezSyntheticGaugeFields2011} or the experimental simulation of quantum-spin Hamiltonians, featuring spin frustration \cite{schmiedQuantumPhasesTrapped2008,feyQuantumCriticalityTwoDimensional2019} and spin-glass dynamics \cite{nathHexagonalPlaquetteSpin2015}.

\begin{acknowledgments}
	The authors thank U. Warring for fruitful discussions and valuable input. This project has received funding from the European Research Council (ERC) under the European Union’s Horizon 2020 research and innovation program (Grant No. 648330).
	D.H. and F.T acknowledge support from the German Science Foundation (DFG) within RTG 2717
	A. M. and T.W. acknowledge additional support from the Georg H. Endress Foundation.
	L. K. acknowledges support from the German Science Foundation (DFG) via the Heisenberg program KA 4215/2-1.
\end{acknowledgments}

\end{document}

\section*{Supplemental Material to "Trapping Ion Coulomb Crystals in an Optical Lattice"}

	\begin{table}[h!] 
				\begin{tabular}{|c|c|c|c|c|c|}
					\hline
					\hline
					\vrule width 0pt height 2.2ex
						$P_{\text{in}}(W)$ & N & $f_{\text{rad,exp}} $(kHz) & $f_{\text{rad,num}}$(kHz)& $f_{\text{ax,exp}}$(MHz) & $f_\text{ax,num}$(MHz) \\
						     3.30(15)      & 1 &                      &                        & 2.495(50) ($\parallel$)           & 3.07(17) ($\parallel$)            \\
						     5.04(15)      & 1 & 32.8(10) (s)      & 35.5(10) (s)          &                     &                     \\
						                   &  & 47.1(10) ($\perp$)       & 50.1(59) ($\perp$)          &                     &                     \\
						                   &   & 60.1(10) ($\parallel$)       & 71.1(82) ($\parallel$)          &                     &                     \\
						     6.30(15)      &   1& 40.0(10) (s)       & 39.9(10) (s)           & 3.220(50) ($\parallel$)    & 4.32(23) ($\parallel$)           \\
						                   &   & 51.1(10) ($\perp$)       & 56.1(61) ($\perp$)           &     &                     \\
						                   &   & 66.8(10) ($\parallel$)      & 79.5(86) ($\parallel$)          &                     &                     \\
						                   &   2&                      &                        &  3.280(50) ($\parallel$)                   &\\
					\hline
					\hline
					\end{tabular}
	\caption{\label{tab:results} Summary of experimentally (exp) and numerically (num) determined  values for the radial (r) and axial (ax) resonance frequencies at different input powers $P_\text{in}$.
		  Values are given for the single-beam (s), lin$\perp$lin ($\perp$) and lin$\parallel$lin ($\parallel$) configuration. N denotes the number of ions.}
	\end{table}